\documentclass[conference]{IEEEtran}
\IEEEoverridecommandlockouts
\usepackage{cite}
\usepackage{amsmath,amssymb,amsfonts}
\usepackage{algorithmic}
\usepackage{graphicx}
\usepackage{textcomp}
\usepackage{xcolor}
\usepackage{url}
\usepackage{hyperref} 
\def\BibTeX{{\rm B\kern-.05em{\sc i\kern-.025em b}\kern-.08em
    T\kern-.1667em\lower.7ex\hbox{E}\kern-.125emX}}

\begin{document}

\title{Correlation of Fr\'echet Audio Distance With Human Perception of Environmental Audio Is Embedding Dependant}

\author{\IEEEauthorblockN{1\textsuperscript{st} Modan Tailleur}
\IEEEauthorblockA{\textit{École Centrale Nantes} \\
\textit{CNRS, LS2N, UMR 6004}\\
F-44000 Nantes, France \\
modan.tailleur@ls2n.fr}
\and
\IEEEauthorblockN{1\textsuperscript{st} Junwon Lee}
\IEEEauthorblockA{
Gaudio Lab, Inc. / KAIST\\
Seoul / Daejeon, South Korea \\
junwon.lee@gaudiolab.com}
\and
\IEEEauthorblockN{2\textsuperscript{nd} Mathieu Lagrange}\IEEEauthorblockA{\textit{École Centrale Nantes} \\
\textit{CNRS, LS2N, UMR 6004}\\
F-44000 Nantes, France \\
mathieu.lagrange@ls2n.fr}
\and
\IEEEauthorblockN{3\textsuperscript{rd} Keunwoo Choi}
\IEEEauthorblockA{
\textit{Gaudio Lab, Inc.}\\
Seoul, South Korea \\
keunwoo@gaudiolab.com}
\and
\IEEEauthorblockN{4\textsuperscript{th} Laurie M. Heller}
\IEEEauthorblockA{\textit{Department of Psychology} \\
\textit{Carnegie Mellon University}\\
Pittsburgh, PA, U.S. \\
hellerl@andrew.cmu.edu}
\and
\IEEEauthorblockN{5\textsuperscript{th} Keisuke Imoto}
\IEEEauthorblockA{\textit{Doshisha University}\\
Kyoto, Japan \\
keisuke.imoto@ieee.org}
\and
\IEEEauthorblockN{6\textsuperscript{th} Yuki Okamoto}
\textit{Ritsumeikan University}\\
Kusatsu, Japan \\
y-okamoto@ieee.org}

\maketitle

\begin{abstract}
This paper explores whether considering alternative domain-specific embeddings to calculate the Fr\'echet Audio Distance (FAD) metric can help the FAD to correlate better with perceptual ratings of environmental sounds. We used embeddings from VGGish, PANNs, MS-CLAP, L-CLAP, and MERT, which are tailored for either music or environmental sound evaluation. The FAD scores were calculated for sounds from the DCASE 2023 Task 7 dataset. Using perceptual data from the same task, we find that PANNs-WGM-LogMel produces the best correlation between FAD scores and perceptual ratings of both audio quality and perceived fit with a Spearman correlation higher than 0.5. We also find that music-specific embeddings resulted in significantly lower results. Interestingly, VGGish, the embedding used for the original Fr\'echet calculation, yielded a correlation below 0.1. These results underscore the critical importance of the choice of embedding for the FAD metric design.
\end{abstract}

\begin{IEEEkeywords}
Environmental Sound Synthesis, Objective Audio Quality, Neural Audio Embeddings, Evaluation Metrics
\end{IEEEkeywords}

\section{Introduction}

Generative audio synthesis has become a popular research topic in which deep neural nets are typically driven by textual prompts\cite{oord2016wavenet, liu2023audioldm, kreuk2022audiogen, ghosal2023tango, huang2023make}. Those systems must be evaluated on high-level perceptual features such as audio quality and alignment with categories for meaningful comparisons. However, evaluating synthetic audio through perceptual evaluation remains a cumbersome process, despite its validity.


To address this challenge, various metrics have been developed for use in prototyping and large-scale quality assessment \cite{elizalde2023clap, liu2023audioldm, huang2022mulan}. Among these metrics, the Fr\'echet Audio Distance (FAD) \cite{kilgour2018fr} is widely used. FAD compares the distribution of a reference set with that of synthetic audio using VGGish embeddings \cite{hershey2017cnn}.

Recent work \cite{gui2023adapting} demonstrates that considering alternative embeddings that have been trained on music data is a relevant approach to better assess the quality of music generation systems.

Similarly, in this paper, we investigate whether changing the embeddings can lead to an increased correlation of the FAD with human perceptual ratings of both audio quality and perceived fit to categories of environmental sounds (i.e., general audio excluding speech and music). Given the fact that VGGish embeddings have been trained on environmental audio, we expected that VGGish embeddings would perform well, but our findings show the opposite. VGGish embeddings report low correlations, as do the embeddings trained on music data. Fortunately, we find that more recent embeddings specifically trained on environmental audio are quite effective, suggesting that the choice of the embedding is a crucial part of FAD metric design.  

The rest of the paper is organized as follows. Section \ref{sec:2} delves into related works, followed by the presentation of the selected embeddings in Section \ref{sec:3}. Section \ref{sec:4} presents the experiments and Section \ref{sec:5} the correlation results between the FAD scores when considering several state-of-the-art neural audio embeddings and perceptual evaluations using the DCASE Task 7 2023 dataset\cite{Choi2023}. The code corresponding to this study is made publicly available.\footnote{Code repository: \url{https://github.com/mathieulagrange/dcaseFadEmbedding }}



\section{Related work \label{sec:2}}

The Fr\'echet Audio Distance (FAD) \cite{kilgour2018fr} has been proposed as an adaptation of the Fr\'echet Inception Distance (FID) \cite{heusel2017gans} for audio quality assessment. FID and FAD compare the distribution of two datasets in a given embedding space. VGGish\cite{hershey2017vggish} was originally proposed as the feature extractor for FAD. To evaluate a synthesis model, a set of desired audio serves as the reference. Firstly, two multivariate Gaussian distributions, which have the same means and covariances as the embedding sets, are considered. Then, the Fr\'echet distance between the two distributions $r$ and $t$ is calculated as follows: 

\begin{equation}\label{eq:FAD}
FAD(r, t) = \left\| \mu_r - \mu_t \right\|_2 + \text{tr} \left( \Sigma_r + \Sigma_t - 2\sqrt{\Sigma_r \Sigma_t} \right)
\end{equation}
where $\mu_x$ and $\Sigma_x$ are respectively the mean and covariance matrix of a given distribution $x$. The FAD calculation compares the two datasets in terms of fit to domain with the comparison of means, but also in terms of diversity by including a form of covariance comparison in the equation. A low FAD score thus indicates that the two datasets contain similar sound sources, and a similar diversity. If the reference dataset can be considered of high audio quality it is generally assumed that a low FAD distance implies that the evaluated dataset is also of good audio quality.

Whether it be for audio or image evaluation, using Fr\'echet distance has a few drawbacks. Trying to match an embedding space can lead to models of very different quality having similar Fr\'echet distance scores, as one could be fitting the embedding space more accurately without improving the perceptual quality\cite{kynkaanniemi2022role}. The majority of embeddings used are trained representations, implying a potentially strong dependency on the dataset and task used for training. Consequently, for accurate evaluation using Fr\'echet Distance, the dataset under assessment must exhibit similarity to the training set of the embedding. This emphasizes the potentially significant influence of the embedding choice on the Fr\'echet Distance calculation process.

For image generation evaluation, Kynkäänniemi, Tuomas, et. al.\cite{kynkaanniemi2022role} showed that simply matching the top-N classifications histogram between the reference and generated set improved the FID score without further improvement in generative model. This indicates the high dependency of FID on ImageNet\cite{deng2009imagenet}, which is used to train the Inception embedding. The high dependency on the choice of embedding may be much worse in audio domain. ImageNet dataset has carefully tailored 1k classes to cover the wide range of in-the-wild images. Conversely, VGGish was trained to classify only 3k labels, that are not even necessarily relevant to sound. This may limit the generalizability of audio embeddings, depending on the training data size and task.

Gui et al. \cite{gui2023adapting} explored the limitations of FAD in music generation, as VGGish-based FAD struggles to accurately predict the perceptual features of generated musical audio.
The authors investigated different embeddings and found that VGGish yields notably poor FAD scores compared to alternative representations. Although they have shown that embeddings such as CLAP \cite{elizalde2023clap, wu_large-scale_2023} are more suitable for music generation, their effectiveness in improving the FAD metric for Environmental audio generation remains to be evaluated. 





\section{Embeddings \label{sec:3}}


Experimental details on all embeddings examined are presented in Table \ref{tab:embedding-models}. Our objective with this selection is to investigate whether domain-specific embeddings significantly influence the relevance of the Fr\'echet Audio Distance (FAD) metric. We consider VGGish \cite{hershey_cnn_2017} as our baseline, as it has a proven record of use for FAD calculation. Next, we consider MERT \cite{li_mert_2023} as a recent embedding primarily trained on music data, as well as a CLAP model trained specifically on voice and music. Given that those latter embeddings are not trained on environmental audio, we hypothesize that they should perform poorly. Other CLAP models and PANNs\cite{kong2020panns} models, on the other end, are trained using environmental audio. As CLAP models are trained using partly PANN architectures, or at least have considered PANNs in their framework in their evaluation protocol, they are expected to outperform PANNs.

\subsection{VGGish}

VGGish \cite{hershey_cnn_2017} is an audio classifier trained on a subset of a large audio dataset extracted from YouTube videos called YouTube-100M, which contains 350,000h of audio data with video-level class labels. YouTube-100M covers a wide range of audio content, spanning everyday sounds, sound effects, and music, captured in diverse real-world scenarios. It's worth noting that within this dataset the video-level classes may not necessarily be directly related to the audio content, as a source can be present in a video without generating any sound. VGGish uses log-mel spectrograms with 64 frequency bins and 10-ms hops as input, and it has about 70M parameters. Following the methodology outlined by Gui~et~al. \cite{gui2023adapting}, we employ the VGGish model to process 1-second audio segments, with 50\% overlap. 



\begin{table}[t]
    \caption{Description of Embedding Models. The size of the Receptive Field (RF) is the maximal duration of audio considered by the model to compute the embedding.}

    \centering
    \begin{tabular}{|l|r|r|r|r|r|}
        \hline
        \textbf{} &  \textbf{Model} & \textbf{Audio } & \textbf{Embedding} & \multicolumn{1}{c|}{\textbf{RF}}  \\
        & \multicolumn{1}{c|}{\textbf{Size}} & \multicolumn{1}{c|}{\textbf{Rate}} & \textbf{Size / Rate} & \textbf{Size}  \\
        \hline
        VGGish \cite{hershey_cnn_2017}& 72M & 16 kHz & 128 / 1 Hz  & 1 s   \\
        MERT \cite{li_mert_2023} & 72M & 24 kHz & 768 / 76 Hz & 5 s  \\
        MS-CLAP \cite{elizalde_natural_2023}& 158M  & 44 kHz & 1024 / 1 Hz & 7 s  \\
        L-CLAP \cite{wu_large-scale_2023} & 158M & 48 kHz & 512 / 1 Hz & 10 s  \\
        PANNs-CNN14-16k \cite{kong2020panns} & 80M & 16 kHz & 2048 / .1 Hz & 10 s \\
        PANNs-CNN14-32k \cite{kong2020panns} & 80M & 32 kHz & 2048 / .1 Hz & 10 s  \\
        PANNs-WGM-Logmel \cite{kong2020panns} & 80M & 32 kHz & 2048 / .1 Hz & 10 s  \\
        \hline
    \end{tabular}
    \label{tab:embedding-models}
\end{table}

\subsection{MERT}

The Music undERstanding model with large-scale self-supervised
Training (MERT) \cite{li_mert_2023} generates embeddings learned with teacher-student methods, using a combination of teachers including an acoustic teacher based on Residual Vector Quantization - Variational AutoEncoder (RVQ-VAE) and a musical teacher based on the Constant-Q Transform (CQT). The student model is a BERT-style transformer encoder. The MERT models are specialized in music, being trained on an in-house private music dataset comprising 160k hours of audio data. Among the several available models, we chose the one with 95M parameters (MERT-95M). 

\subsection{PANNs}

PANNs \cite{kong2020panns} are classifiers trained on AudioSet\cite{gemmeke2017audioset}. The dataset originated from YouTube videos and is around 5,000h long, with 527 different general sound classes. Unlike YouTube-100M, AudioSet employs human labeling at the audio level. The majority of these models leverage log-Mel spectrograms featuring 64 mel bins and 10-ms hops as input. Released in various sizes and trained at different sample rates, these models offer versatility in application. Among them, the CNN14 model emerges as the most commonly utilized variant, with 16kHz (PANN-CNN14-16kHz) or 32kHz (PANN-CNN14-32kHz) sampling rates.  However, the model that leads to the best accuracy, according to their paper, is the Wavegram-Logmel-CNN (PANN-WGM-LogMel). This model uses the audio waveform as input, which is transformed into a learned spectro-temporal representation along with a Mel-spectrogram. Subsequently, both representations are fed into the rest of the network. PANN-CNN14 and PANN-WGM-LogMel models contain 80M parameters.



\subsection{MS-CLAP}

The embedding Contrastive Language-Audio Pretraining (CLAP) \cite{elizalde2023clap} is trained to learn multimodal representations, using both an audio and a text encoder. Symmetric cross-entropy loss was exploited for language and audio cross-modal contrastive learning. It uses PANN-CNN14-32kHz for audio encoding and BERT for text encoding. It is trained on 128,000 audio/caption of FSD50k, ClothoV2, AudioCaps and MACS, representing about 250h of audio. For this study, we use the CLAP model released in September 2023 called ``sep 23". 


\subsection{L-CLAP}

LAION has also trained a CLAP architecture \cite{wu_large-scale_2023} similar to the one used for MS-CLAP. They trained their best model using HTS-AT \cite{chen2022hts} as the audio encoder and RoBERTa \cite{liu2019roberta} as the text encoder. They proved that training their model on LAION-Audio-630k and AudioSet with keyword-to-caption augmentation significantly improves the performances of their CLAP model. LAION-Audio-630k comprises a diverse range of 4,000 hours of audio recordings depicting human activities, natural sounds, and audio effects, sourced from eight publicly available websites. In total, their model for environmental audio called ``630k-audioset-best” (L-CLAP-audio) is reported to be trained on about 10,000h of audio. They also released models specialized in music, which are trained on music and speech from their data collection. We chose the model “music audioset epoch 15 esc 90.14” (L-CLAP-mus) for comparison.

\section{Experiments \label{sec:4}}

\subsection{Data}

In this study, we leverage the DCASE 2023 Challenge Task 7 dataset\cite{Choi2023}, encompassing 700 sound excerpts of 7 different categories: \textit{dog bark, footstep, gunshot, keyboard, moving motor vehicle, rain, and sneeze/cough}. Each sound excerpt is a mono 16-bit 22,050 Hz 4-second audio sourced from three distinct datasets: UrbanSound8K\cite{salamon2014urban8k}, FSD50K\cite{fonseca2021fsd50k}, and BBC Sound Effects\footnote{\url{https://sound-effects.bbcrewind.co.uk/}}. To ensure high relevance, diversity, and clarity, the challenge organizers manually selected and validated the excerpts.

Supplementary to this dataset are the audio generated using the baseline and 8 submitted algorithms, with each contributing an additional set of 700 sound excerpts synthesized by their respective systems. The duration of the whole dataset (recorded and synthesized) is about 8h.  
The 8 systems from the participants were top-ranked in terms of FAD score. Twenty sounds from each system underwent perceptual evaluation for both audio quality and category fit by 91 raters (47 hrs). Each category within each system has been evaluated, resulting in a total of 63 evaluations for each criterion. The correlations shown in section \ref{sec:5} are thus based on those 20 evaluated audios per category and per system.

\subsection{Uncertainty estimation}

To assess the uncertainty associated with each Spearman correlation calculation, we introduce Gaussian noise with a standard deviation of 1 to the perceptual evaluation scores. Subsequently, we repeat this process 100 times to generate 100 different noisy sets, each containing 63 perceptual evaluations for both category fit and audio quality.

By computing the mean and standard deviation of the Spearman correlation coefficients across these 100 noisy sets, we obtain estimates of the variability and uncertainty inherent in the correlation calculations. 

\section{Results \label{sec:5}}

Given the FAD definition given in Eq. \ref{eq:FAD}, lower FAD scores indicate a high similarity between datasets. Consequently, a reliable FAD metric should demonstrate an inverse correlation with high perceptual quality when computed between a reference set and various generated sets. To simplify the presentation of results, we use the inverse of the FAD ($\text{FAD}^{-1}$) so that a high quality $\text{FAD}^{-1}$ metric should achieve a high positive correlation with perceptual attributes that are higher when better.

\begin{figure}[t]
\begin{center}
\includegraphics[width=\linewidth]{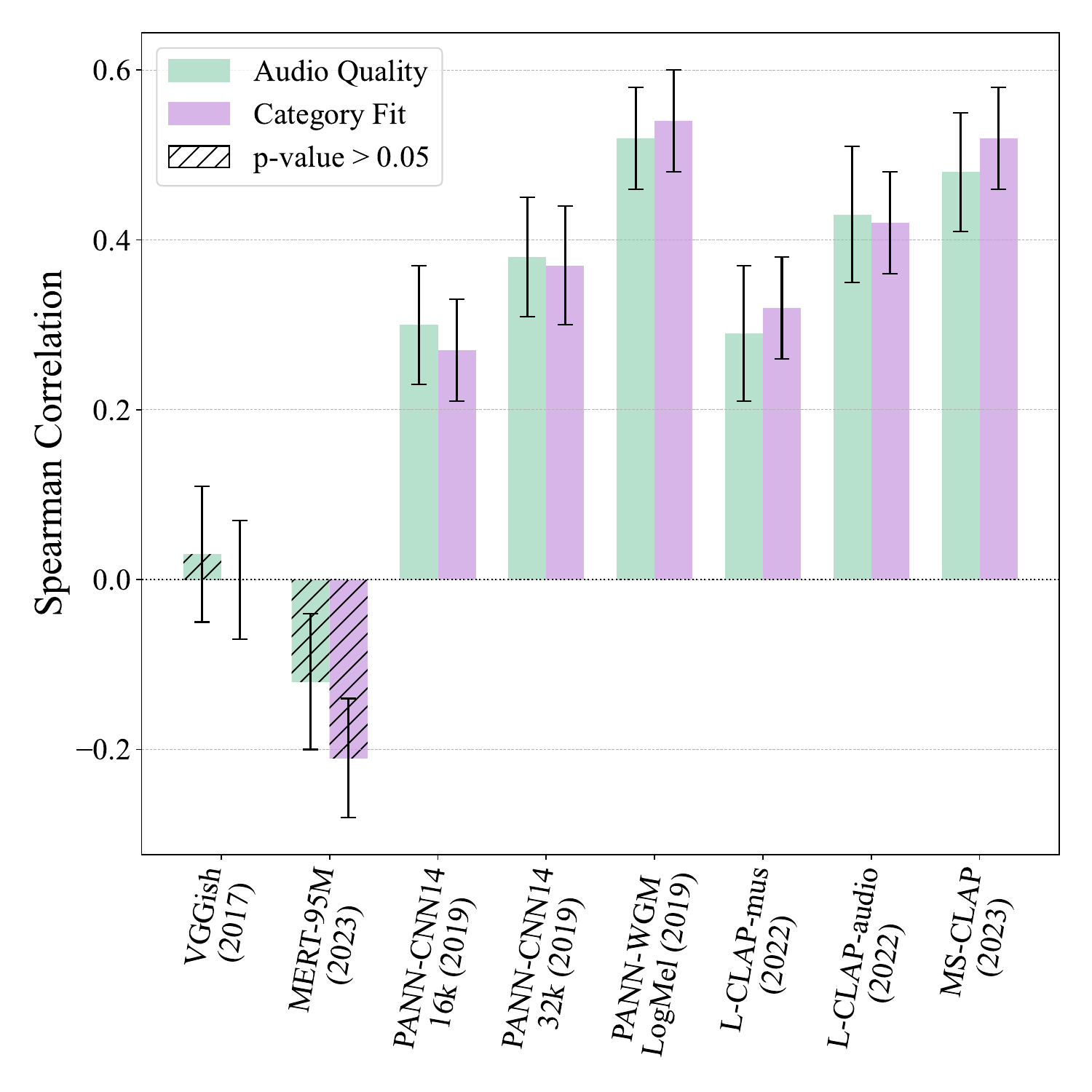}
\end{center}
\vspace{-0.5cm}
\caption{Spearman correlation coefficient ($n=63$) between $\text{FAD}^{-1}$. and perceptual evaluation of audio quality and category fit for different embeddings. Error bars display  standard deviation.}
\vspace{-0.4cm}
\label{fig:plot_bar_graph_correlation}
\end{figure}

\subsection{Overall Correlation}

As shown in Figure \ref{fig:plot_bar_graph_correlation}, the PANNs-WGM-LogMel FAD and the MS-CLAP FAD demonstrate significantly high correlations with both category fit and audio quality, with PANN-WGM-LogMel being significantly higher than MS-CLAP FAD. In contrast, both the MERT-95M FAD and the VGGish FAD demonstrate very weak correlation with perceptual evaluation. Additionally, the L-CLAP models perform less effectively compared to MS-CLAP and PANN-WGM-LogMel, with L-CLAP-audio showing a better correlation score than L-CLAP-mus.

\subsection{Per-category Correlation}

\begin{figure}[t]
\begin{center}
\includegraphics[width=\linewidth]{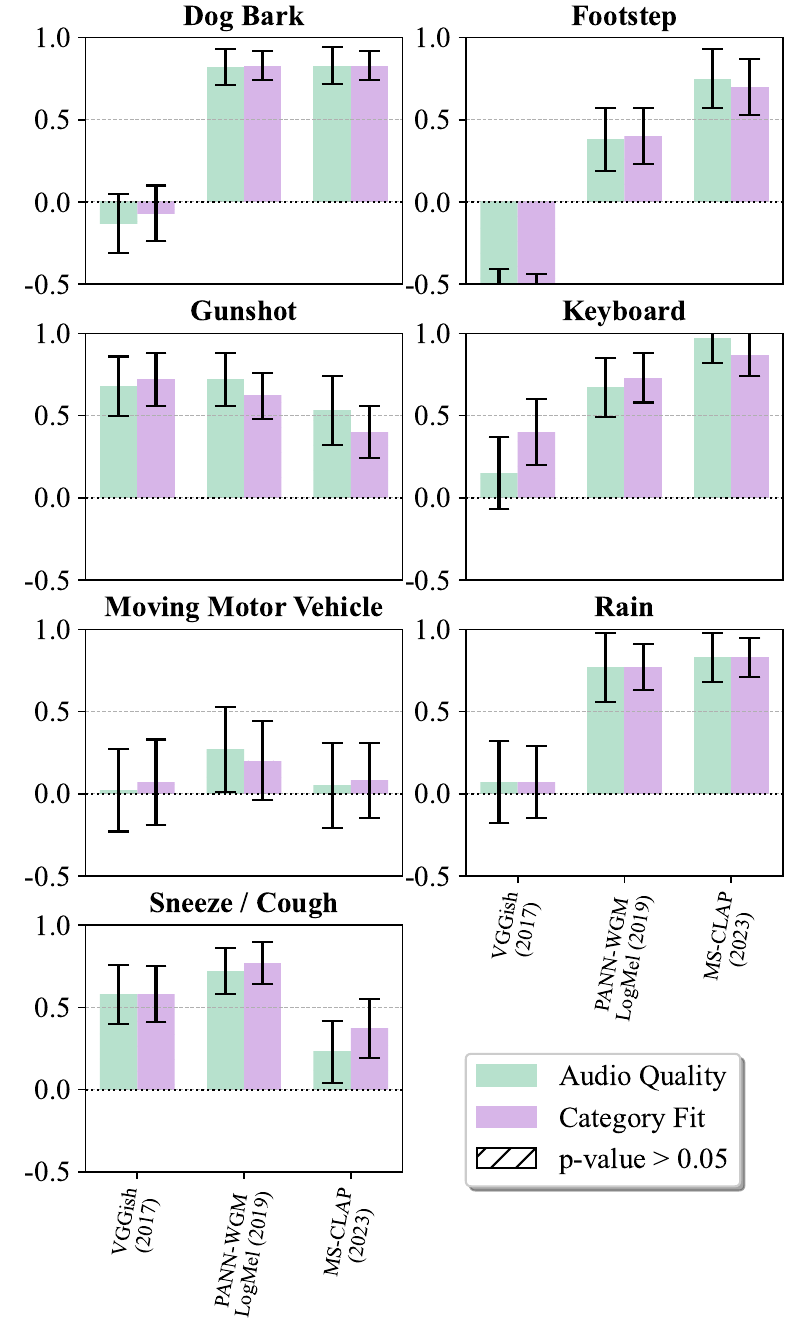}
\end{center}
\vspace{-0.5cm}
\caption{Spearman correlation coefficient ($n=9$) between $\text{FAD}^{-1}$. and perceptual evaluation of audio quality and category fit for VGGish, PANNs CNN14 Wavegram Logmel and CLAP.}
\vspace{-0.4cm}
\label{fig:plot_bar_graph_correlation_all_categories}
\end{figure}

In Figure \ref{fig:plot_bar_graph_correlation_all_categories}, we present the per-category correlation results. Interestingly, substantial variability in correlation is observed across different categories. PANNs-WGM-LogMel displays greater stability across categories, while performing better than the other embeddings on half of the categories. VGGish demonstrates good performance in categories such as sneeze/cough and gunshot, but performs extremely poorly in every other category. Similarly, CLAP exhibits low correlations in categories such as moving motor vehicle, sneeze/cough, and gunshot, while demonstrating better performance in other categories, competing or beating PANNs-WGM-LogMel in some of them. These results should be treated cautiously though as the Spearman correlation coefficients are calculated with only 9 data points per category. 

\subsection{Influence of Dimensionality}

As shown in Table \ref{tab:embedding-models}, each embedding model varies in size. Considering the findings presented in Figure \ref{fig:plot_bar_graph_correlation}, we want to investigate if the dimensionality of the embedding may bias the performance of the FAD metric. To refute the hypothesis that a higher dimensional embedding may gain an unfair advantage, we conduct a dimensionality reduction of every embedding to match that of VGGish ($n$=128). The dimensionality reduction is performed by projecting the set of embeddings on the 128 Eigenvectors of the highest Eigenvalues using Principal Component Analysis (PCA).

It is observed that diminishing the size of the embeddings had minimal impact on the correlations with $\text{FAD}^{-1}$, decreasing them by approximately 0.01 for each embedding. This tends to show us that the size of the embedding does not significantly influence the performance of the FAD metric.


\subsection{FAD-based Category Mapping}

Little is known about how the FAD may lead to a similarity graph that is linked to the perceptual attributes of sounds. To examine this, we grouped the 7 sound categories of the 700 audios of the evaluation set of DCASE Task 7 2023 dataset into 3 meta-categories: Impact (Footsteps, Gunshot, Keyboard), Living (Dog Bark, Sneeze/Cough), and Texture (Moving Motor Vehicle, Rain). A satisfactory embedding should result in the FAD grouping similar categories together while separating highly dissimilar categories. Given that the 7 sound classes themselves are highly distinct, they should also be quite separated from each other in the projection.

\begin{figure*}[!htb]
\begin{center}
\includegraphics[width=\linewidth]{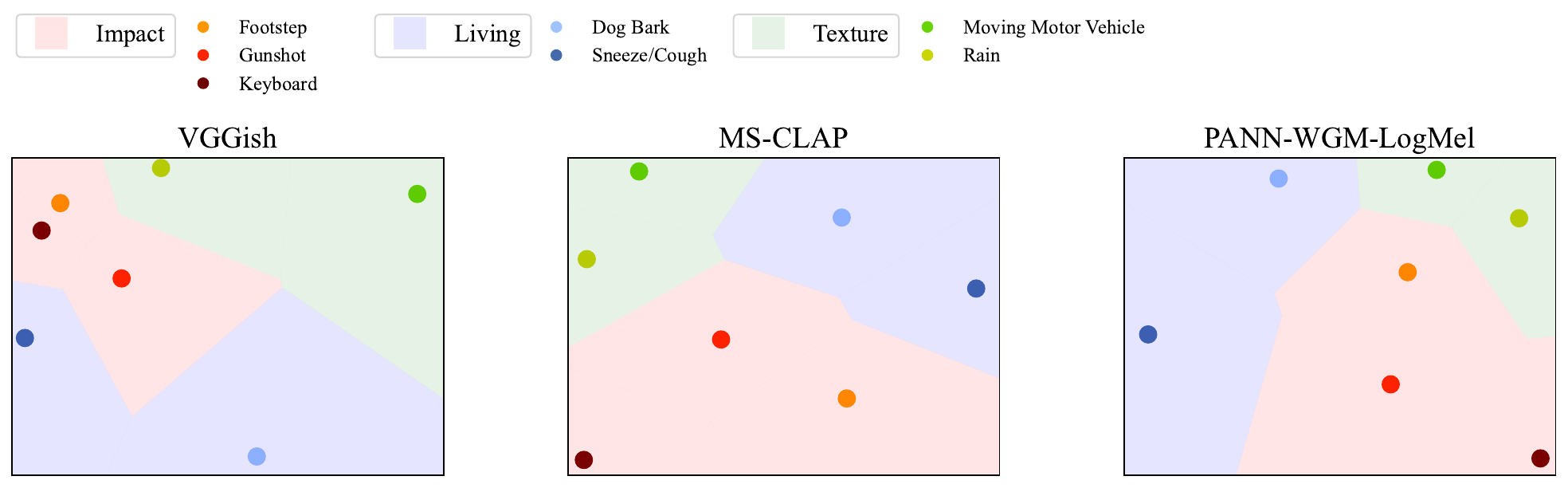}
\end{center}
\vspace{-0.5cm}
\caption{2D Projection of inter-category FAD similarity matrix using Multi-dimensional scaling (MDS) on DCASE Task 7 2023 dataset.}
\vspace{-0.4cm}
\label{fig:plot_dcase_isomap}
\end{figure*}

Figure \ref{fig:plot_dcase_isomap} shows a 2D mapping of the similarities of every pair of categories from the DCASE Task 7 2023 dataset, calculated with Multidimensional Scaling (MDS) with FAD using three different embeddings as input. 
We find that the three embeddings successfully group similar categories together, as evidenced by the Voronoi 
separation diagram. Additionally, the split between the different groups is somewhat more pronounced for MS-CLAP and PANN-WGM-LogMel than for VGGish. Furthermore, for MS-CLAP and PANN-WGM-LogMel, all 7 sound sources are further apart from each other, which indicates that these two embeddings are the best at separating the different categories.

\section{Conclusion}

In this paper, we explored the use of alternative embeddings to assess audio quality and alignment with categories in environmental audio, aiming to improve the validity of the Fr\'echet Audio Distance (FAD) metric. We compared several embeddings, including VGGish, PANNs, MS-CLAP, L-CLAP, and MERT, using the DCASE Task 7 2023 dataset. 


We find that there is a strong dependency of the embedding in the FAD metric, which appears to be closely linked to the domain of the embedding training dataset. In fact, music-trained embeddings, such as MERT-95M and CLAP Laion Music, perform less effectively than those trained on environmental audio. Furthermore, while VGGish was trained to classify 3k video-level labels, these labels might not necessarily be relevant to sound, potentially limiting its generalizability for tasks like DCASE TASK 7 2023 systems evaluation. This observation also aligns with findings reported by Gui et al. \cite{gui2023adapting}.

Interestingly, among the PANN models, only PANN-WGM-LogMel consistently outperforms MS-CLAP. One explanation could be that only the PANN-CNN14-32kHz model was evaluated and/or incorporated by MS-CLAP and L-CLAP. However, according to Kong et al. \cite{kong2020panns}, PANN-WGM-LogMel is the best performing model for classification on Audioset. Thus, a CLAP model trained with PANN-WGM-LogMel as an audio classifier could be a relevant improvement.

It is important to note that the performance of the embeddings varies significantly depending upon the category. This indicates that domain knowledge is important for selecting an appropriate embedding for the FAD calculation.

The study suggests that specialized embeddings tailored to specific tasks are, as of today, crucial for the relevance of the FAD metric. Further investigation is recommended, for example, by considering a more diverse number of categories for which perceptual ratings are available.

\bibliographystyle{ieeetr} 
\bibliography{reference}

\end{document}